\documentclass[sigconf]{aamas} 
\renewcommand\footnotetextcopyrightpermission[1]{} 
\usepackage{balance} 
\usepackage{amsmath}
\usepackage{mathtools}
\usepackage{tikz}
\usepackage{dsfont}

\definecolor{myGreen}{RGB}{31, 150, 2}
\definecolor{myBlue}{RGB}{20, 82, 176}
\usepackage{colortbl}

\newcommand{\calU}{\mathcal{U}}
\newcommand{\calI}{\mathcal{I}}
\newcommand{\ind}[1]{\mathds{1}[#1]}
\newcommand{\bftheta}{\pmb{\theta}}
\newcommand{\bfpi}{\pmb{\pi}}
\newcommand{\bfsigma}{\pmb{\sigma}}
\newcommand{\bfmu}{\pmb{\mu}}

\newcommand{\upa}{\textcolor{myGreen}{\uparrow}} 
\newcommand{\downa}{\textcolor{red}{\downarrow}}

\acmSubmissionID{716}

\title[Improving Peer Assessment with Graph Convolutional Networks]{Improving Peer Assessment with Graph Convolutional Networks}

\author{Alireza A. Namanloo}
\affiliation{
  \institution{Ontario Tech University}
  \city{Oshawa}
  \state{Ontario}
  \country{Canada}}
\email{ali@ali-naman.com}

\author{Julie Thorpe}
\affiliation{
  \institution{Ontario Tech University}
  \city{Oshawa}
  \state{Ontario}
  \country{Canada}}
\email{julie.thorpe@ontariotechu.ca}

\author{Amirali Salehi-Abari}
\affiliation{
  \institution{Ontario Tech University}
  \city{Oshawa}
  \state{Ontario}
  \country{Canada}}
\email{abari@ontariotechu.ca}

\begin{abstract}
Peer assessment systems are emerging in many social and multi-agent settings, such as peer grading in large (online) classes, peer review in conferences, peer art evaluation, etc. However, peer assessments might not be as accurate as expert evaluations, thus rendering these systems unreliable. The reliability of peer assessment systems is influenced by various factors such as assessment ability of peers, their strategic assessment behaviors, and the peer assessment setup (e.g., peer evaluating group work or individual work of others). In this work, we first model peer assessment as multi-relational weighted networks that can express a variety of peer assessment setups, plus capture conflicts of interest and strategic behaviors.  Leveraging our peer assessment network model, we introduce a graph convolutional network which can learn assessment patterns and user behaviors to more accurately predict expert evaluations. Our extensive experiments on real and synthetic datasets demonstrate the efficacy of our proposed approach, which outperforms existing peer assessment methods.
\end{abstract}

\keywords{Peer Assessment, Multi-Agent Systems, Graph Neural Network.}

\newcommand{\BibTeX}{\rm B\kern-.05em{\sc i\kern-.025em b}\kern-.08em\TeX}

\setcopyright{none}

\begin{document}

\ccsdesc[500]{Computing methodologies~Multi-agent systems}
\ccsdesc[300]{Computing methodologies~Neural networks}


\pagestyle{fancy}
\fancyhead{}


\maketitle 


\section{Introduction}
Peer assessment systems have emerged as a cost-effective and scalable evaluation mechanism in many multi-agent settings such as peer grading in large (online) classes, peer review in conferences, peer art evaluation, etc. In these systems, peers assess each others' work (e.g., assignments, papers, etc.) in lieu of a set of pre-appointed experts responsible for evaluation (e.g., instructors, teaching assistants, program committee members, etc.). These peer assessment systems not only make the evaluation of thousands of contributions plausible, but also help to deepen peers' understanding \cite{sadler2006impact}, and facilitate peers providing feedback to each other \cite{psenicka2013impact}. These benefits of peer assessment systems have come with some social and technical challenges, which impact their reliability and robustness.

The reliability of peer assessment systems is directly impacted by the accuracy of peers in their assessments. Peers might lack knowledge or motivation to accurately evaluate others, or they might be strategic in their assessments for their own gain. Two classes of approaches are taken to address these challenges. One primarily focuses on designing strategy-proof peer assessment mechanisms, which incentivize peers to accurately assess each other \cite{de2016incentives, staubitz2016improving, IncentivizingLimited2019,miller2005eliciting,jecmen2020mitigating,zarkoob2020report,wright2015mechanical}. The other class of approaches---most relevant to our work---emphasizes learning peer aggregation mechanisms, which aggregate noisy peer assessments for an item (e.g., assignment or paper) as an estimate of its ground-truth valuation (or expert evaluation) \cite{tunedmodel, peerRank14, fang2017rankwithta, wang2019sspa, vancouver}. 


The learning methods for peer assessment aggregation fall into unsupervised \cite{peerRank14,tunedmodel,de2014crowdgrader}, and semi-supervised \cite{fang2017rankwithta,wang2019sspa} approaches based on whether or not a subset of ground-truth labels are used for training in addition to peer assessment data. These models usually possess particular inductive biases such as peer's accuracy in assessment is correlated with his/her item's ground-truth valuations (e.g., the grade of his/her assignment) \cite{tunedmodel, peerRank14, fang2017rankwithta}; or peer's accuracy in an assessment depends on the extent of its agreement with others' assessments or ground-truth valuations \cite{wang2019sspa}. However, these machine learning methods are empirically shown to be only as effective as simple aggregation mechanisms such as averaging \cite{sajjadi}. Moreover, these approaches are not flexible and general enough to accommodate a wide variety of peer assessment modes (e.g., when an individual assesses the group contribution of others or self assessments).  Our focus in this paper is to develop a semi-supervised aggregation mechanism without any specific or restrictive inductive bias, accommodating various modes of peer assessments.    

%
%
%
We first introduce our graph representation model of peer assessment, which we call \emph{social-ownership-assessment network (SOAN)}. Our SOAN model can express a wide variety of peer assessment setups (e.g., self-assessment and peer assessment for both individual or group contributions) and represent conflict-of-interest relations between peers using auxiliary information (e.g., social networks). Leveraging our SOAN model, we then introduce a semi-supervised graph convolution network (GCN) approach, called \emph{GCN-SOAN}, which can learn assessment patterns and behaviors of peers, without any restrictive inductive bias, to predict the ground-truth valuations. We run extensive experiments on real-world and synthetic datasets to evaluate the efficacy of GCN-SOAN.  Our GCN-SOAN outperforms a wide variety of baseline methods (including simple heuristics, semi-supervised, and unsupervised approaches) on the same real-world dataset \cite{sajjadi}, which was shown to be challenging for machine learning approaches. We further analyze the robustness of GCN-SOAN on a wide range of synthetic data, which captures strategic assessment behavior between users, follows the assumptions of competitor baselines, or considers strict and generous graders. In all those datasets, even those that were tailored to the baselines, our GCN-SOAN outperforms others. Our GCN-SOAN approach can be a stand-alone approach or possibly be integrated with some existing mechanisms for incentivizing accurate assessments (e.g., \cite{wright2015mechanical,de2014crowdgrader,zarkoob2020report}).

\section{Related Work}
We review the related work on peer assessment aggregation methods, strategic behavior in peer assessment systems, incentivizing peers for accurate assessment, and node representation learning in graphs. 

\vskip 1.5mm
\noindent \textbf{Peer assessment aggregation methods.} Peer assessment aggregation methods fall into two categories from the machine learning perspective: unsupervised \cite{peerRank14,tunedmodel,vancouver} and semi-supervised \cite{fang2017rankwithta,wang2019sspa}. In an unsupervised approach, the goal is to aggregate peer assessments for an item (e.g., assignment) to estimate its ground-truth valuation (or grade) without using ground-truth labels. Average and Median---the most simple yet effective form of unsupervised aggregation methods---take the average and median of the peer assessments to estimate ground-truth valuations. In addition to these methods, there is a growing body of unsupervised methods which use the grade of each user's submission as a proxy of his/her accuracy in grading others \cite{peerRank14,tunedmodel}. These methods use various techniques such as iterative methods similar to PageRank \cite{peerRank14}, Bayesian inference by sampling \cite{tunedmodel}, or expectation maximization \cite{vancouver}. In semi-supervised aggregation methods, the goal is to find the aggregation method for predicting ground truth valuations while using all peer assessments and a small subset of ground-truth labels as training data. Some models in this category are again based on the assumption that the user's submission's grade should be predictive of his/her grading accuracy \cite{fang2017rankwithta}. However, some others obtain the grading accuracy of graders by comparing their assessments directly and indirectly to ground truth grades of a set of items \cite{wang2019sspa}.

Our work is a semi-supervised peer assessment aggregation method. However, our work is different from other aggregation methods in several ways: (i) it doesn't make any specific inductive assumption (e.g., correlation of grader's accuracy with his/her grades of items); (ii) it can accommodate various modes of assessment ranging from self-evaluation to peer assessment of group work; (iii) our solution benefits from auxiliary information (e.g., social networks) to capture conflict-of-interest relations between peers and correct its prediction.

\vskip 1.5mm
\noindent \textbf{Strategic behaviors and incentivizing accurate assessment.} Strategic manipulation in peer assessment systems are those errors and inaccuracies in peer evaluations or systems, which are intentionally introduced by the users to get a more favorable outcome or assessment \cite{alon2011sum, aziz2016strategyproof, kahng2018ranking, jecmen2020mitigating, huang2017discovery}.  In a peer selection setting, an individual might misreport his/her ranking of other peers or their work to increase the chance of selection of someone's else (e.g., committee selection, award selection, etc.) or that person's work (e.g., conference review) \cite{alon2011sum, aziz2016strategyproof, kahng2018ranking, aziz2019strategyproof, kotturi2020hirepeer, StrategyproofConference2019, stelmakh2020catch}. Even in those settings that users do not compete with each other (e.g., classrooms), users might mutually agree to strategically inflate or deflate grades to reach their desirable outcome or assessment \cite{reily2009two}. Since peer assessment is time-consuming, users might not put the effort to carefully evaluate peers' items and might give them higher/lower evaluation \cite{IncentivizingLimited2019}. To motivate accurate grading behaviors, many solutions are proposed for rewarding graders based on whether their assessments agreed with other peers' assessments or ground-truth grades \cite{wright2015mechanical, wang2018optimal, IncentivizingLimited2019,  zarkoob2020report, wang2019sspa}. These solutions have a close connection to peer prediction models with the focus on designing the mechanisms to motivate users to put effort in truthful disclosure of private information  \cite{miller2005eliciting, radanovic2014incentives, zhang2014elicitability, shnayder2016informed, kong2016putting, shnayder2016informed, agarwal2020peer, dasgupta2013crowdsourced}. 

Our work differentiates from this literature in several ways. Our goal is not to incentivize accurate assessment, nor to detect strategic behaviours. We aim to learn semi-supervised aggregation functions using graph neural networks, which learns to aggregate inaccurate assessments (even strategic assessments) to predict the ground-truth valuations. Our solution can be stand-alone or possibly be integrated with some existing mechanisms for incentivising accurate assessments (e.g, \cite{wright2015mechanical,de2014crowdgrader,zarkoob2020report}).

\vskip 2mm
\noindent \textbf{Node representation learning in graphs.}
The goal of node representation learning in a graph is to encode each node into a low-dimensional embedding space such that similar nodes in the graph are embedded close to each other.  Node embedding approaches can be categorized into semi-supervised \cite{GCN17,GAT} or unsupervised \cite{perozzi2014deepwalk,grover2016node2vec}. DeepWalk \cite{perozzi2014deepwalk} and Node2Vec \cite{grover2016node2vec}, two popular examples of unsupervised methods, use random walks starting from each node to generate a set of walks (or node's neighborhood). These random walks are then deployed analogous to sentences in Word2Vec \cite{word2vec} to extract the representation of each node, capturing both local and global information of each node in the graph. Graph Neural Networks (GNNs)  \cite{gilmer2017neural} learn node embeddings by iteratively aggregating nodes' information from its local neighborhood. This iterative aggregation allows the node to gradually aggregate more information from farther nodes in the graph. The first iteration only captures the one-hop neighborhood; but after $k$ iterations, the information from the k-hop neighborhood is encoded to node embeddings. GNNs differentiate by their aggregation functions. Some notable examples include Graph Convolutional Network (GCN) \cite{GCN17}, GraphSAGE \cite{graphsage}, and Graph Attention Networks \cite{GAT}. The GCN is one of the most popular semi-supervised GNNs that uses a symmetric normalization technique coupled with self-loops to aggregate and update node embeddings. Our work is built upon GCN.

\section{Proposed Models and Algorithms}\label{sec:GCN-SOAN}
Our goal is to predict the ground-truth assessments (e.g., expert assessments of educational or professional work) from noisy peer assessments. We here first discuss our proposed graph representation model, \emph{social-ownership-assessment network (SOAN)}, for capturing the peer grading patterns and behavior.\footnote{SOAN reads as ``swan.''} Then, we present a modified graph convolution network (GCN) approach, which leverages our SOAN model, to predict the ground-truth assessments.  We call this approach \emph{GCN-SOAN}.  
\subsection{Social-Ownership-Assessment Model}
We assume that a set of $n$ users $\calU$ (e.g., students or scholars) can assess a set of $m$ items $\calI$ (e.g., a set of educational, professional, or intellectual work). The examples cover various applications ranging from peer grading in classrooms to peer reviewing scientific papers, professional work, or research grant applications. We also consider each item $i \in \calI$ possesses a (possibly unknown, but verifiable) ground-truth value $v_i \in \mathbb{R}^+$ (e.g., staff grade for a course work, or expert evaluation of intellectual or professional work). 

\begin{figure*}[tb]
    \centering
    \begin{tabular}{cccc}
      \includegraphics[width=0.22\textwidth]{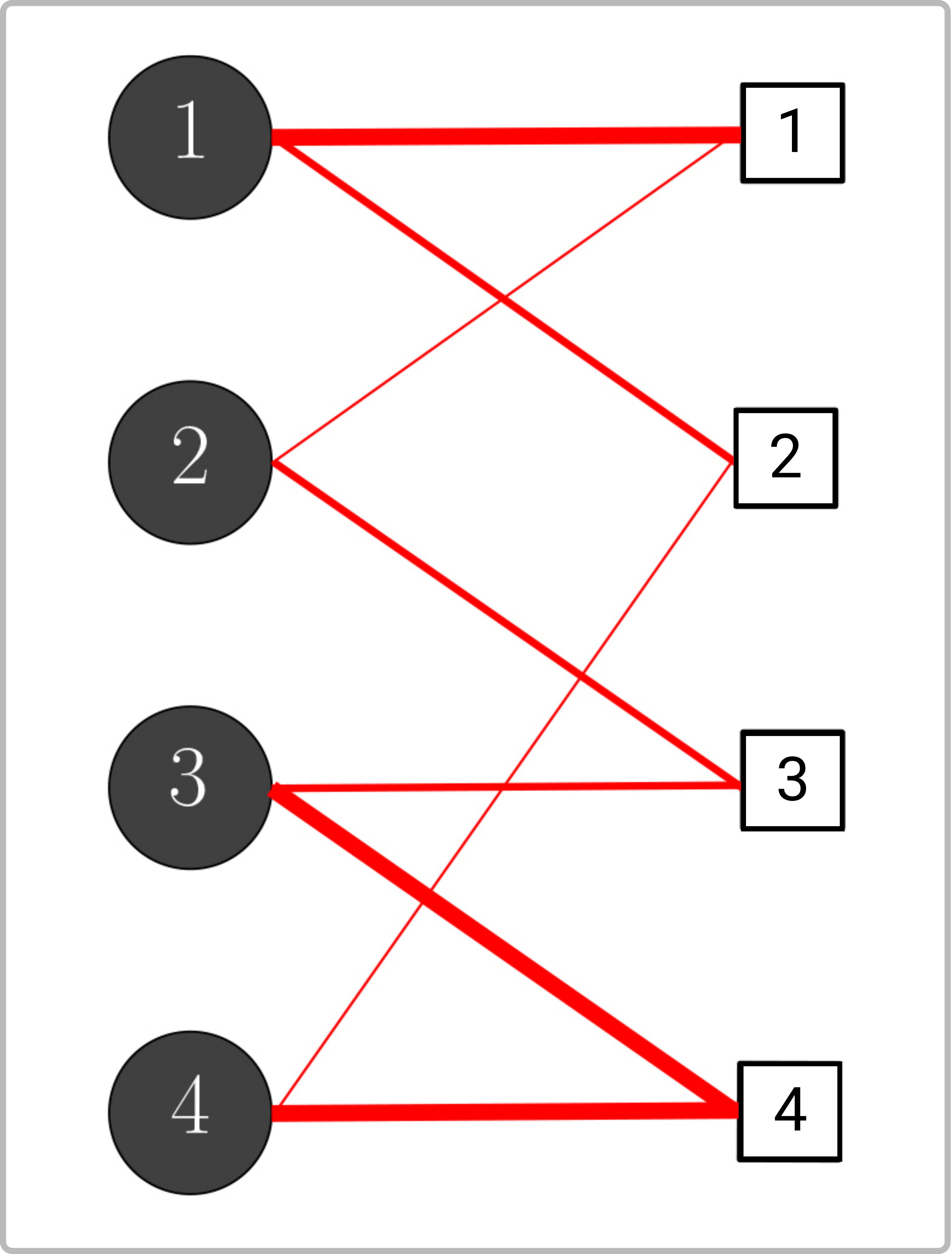}  &
      \includegraphics[width=0.22\textwidth]{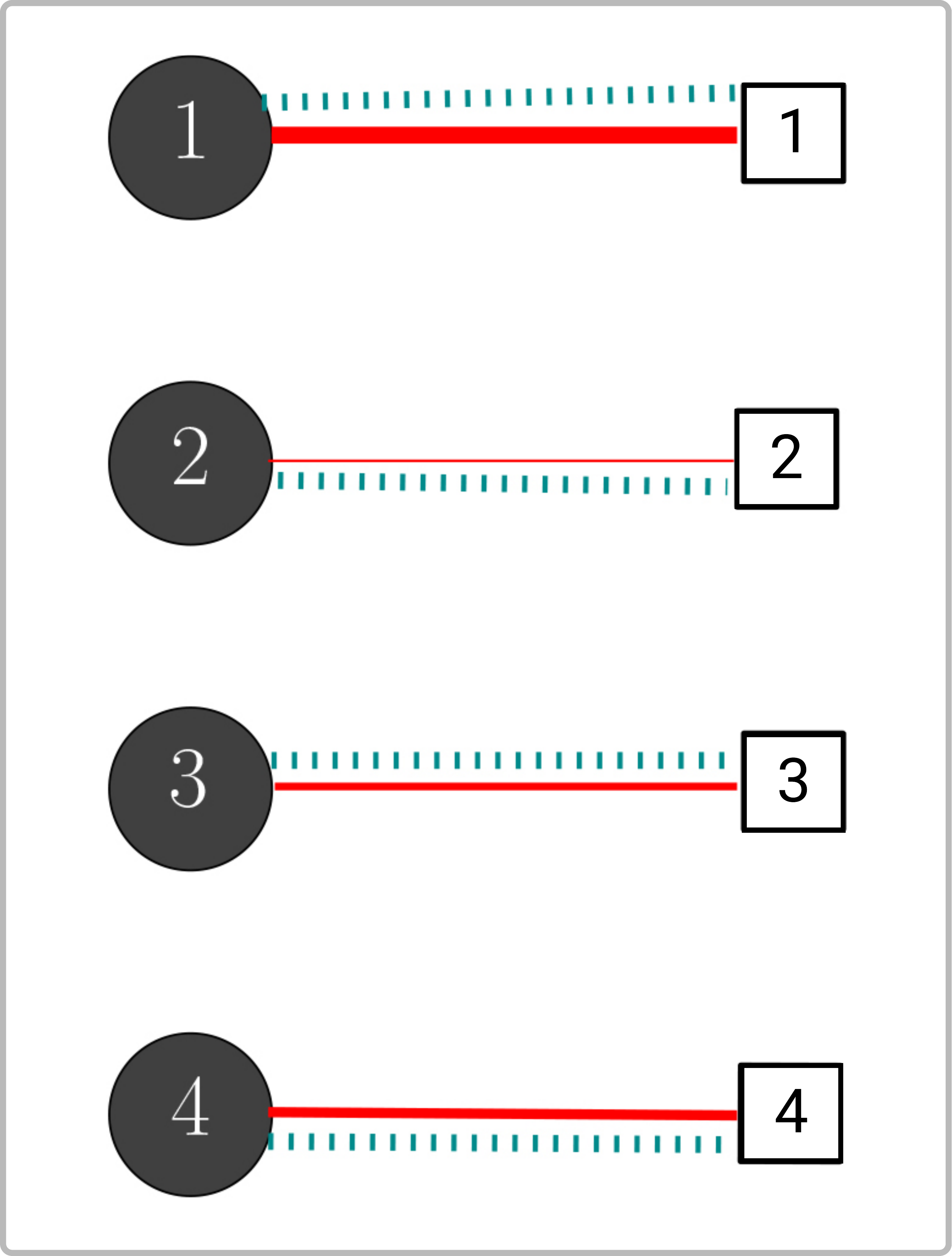}  &
      \includegraphics[width=0.22\textwidth]{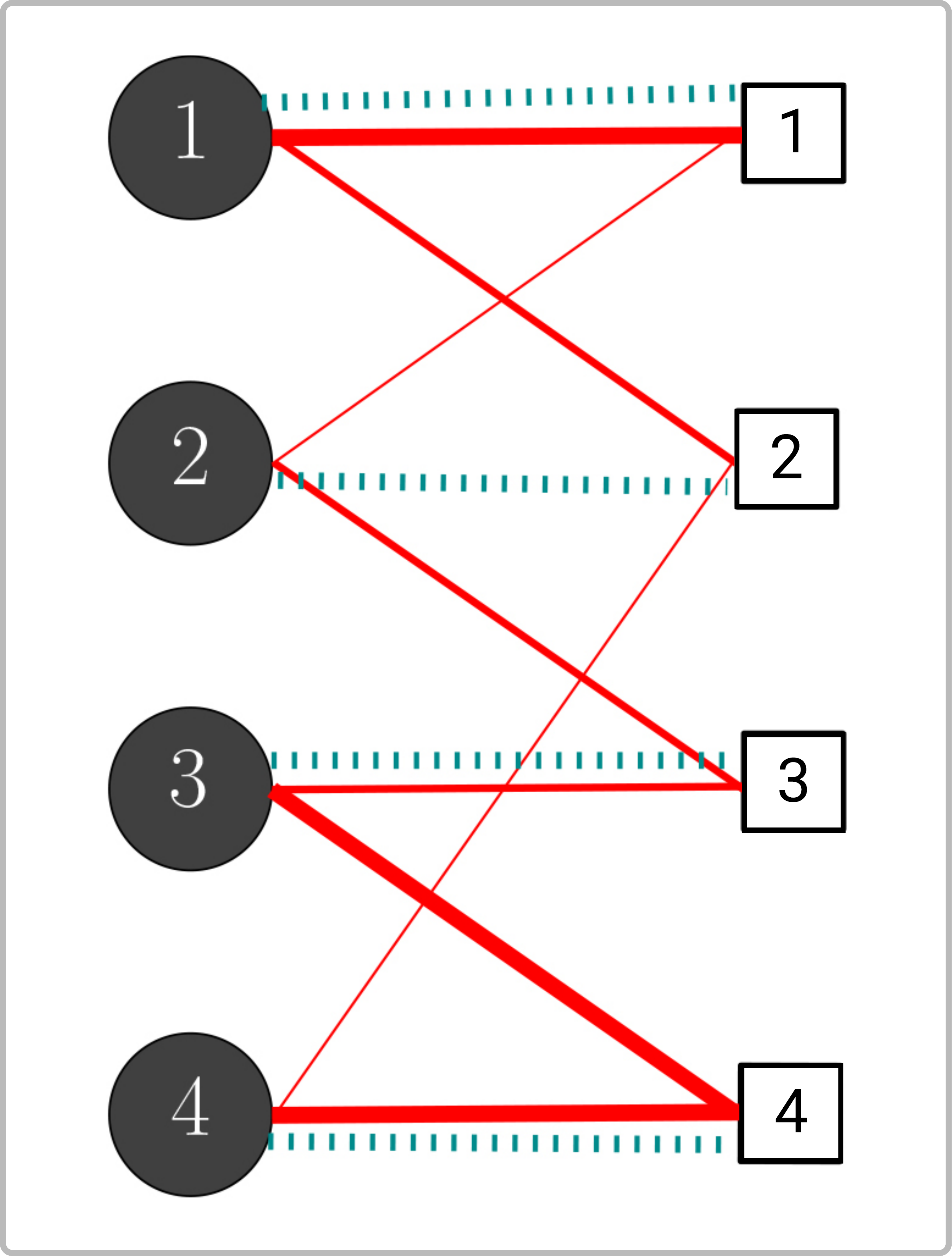} &
      \includegraphics[width=0.22\textwidth]{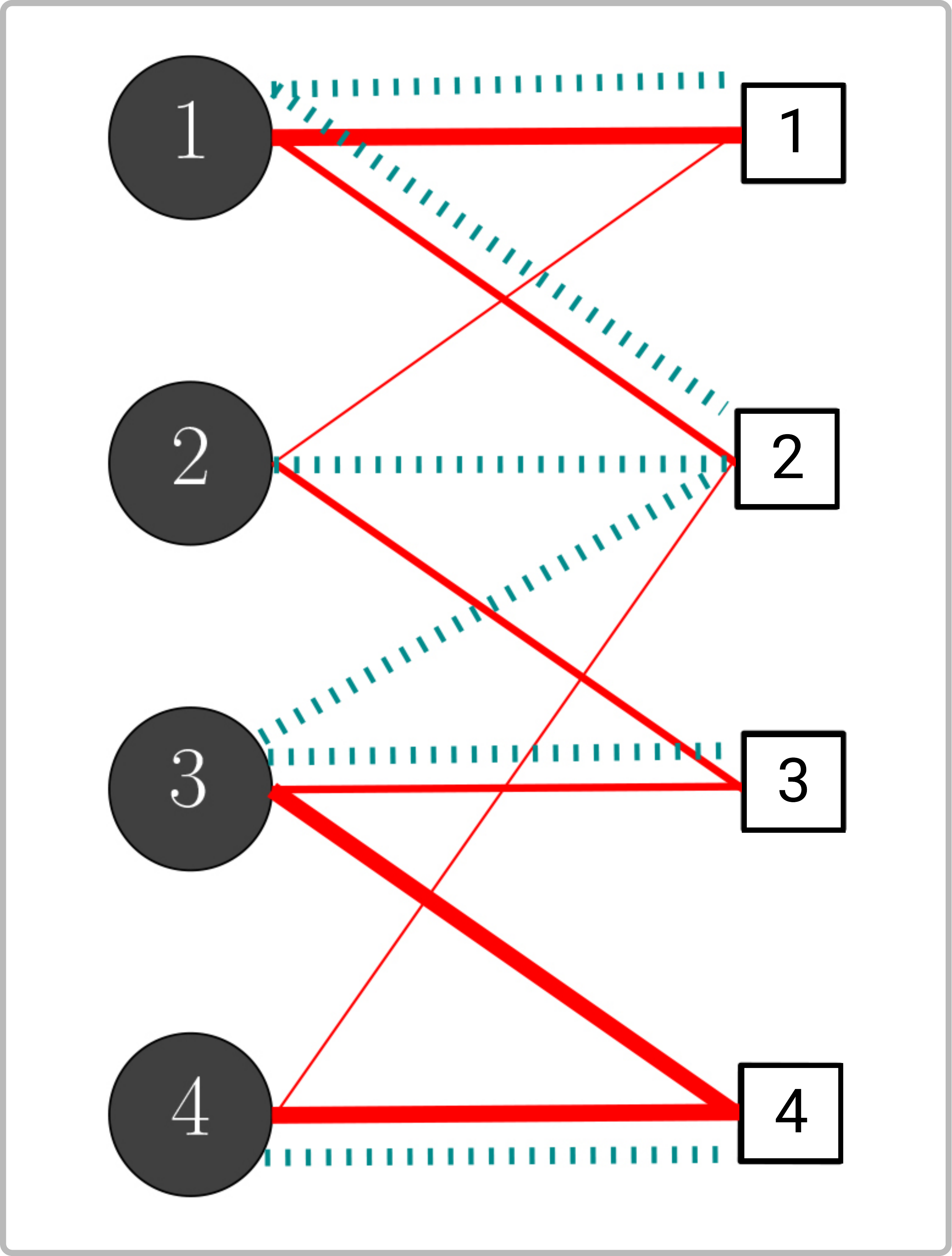}\\
      (a) \small Assessment & (b) \small Self Assessment 
      &(c) \small Assessment, Solo Ownership 
      &(d) \small Assessment, Group Ownership\\ 
      \includegraphics[width=0.22\textwidth]{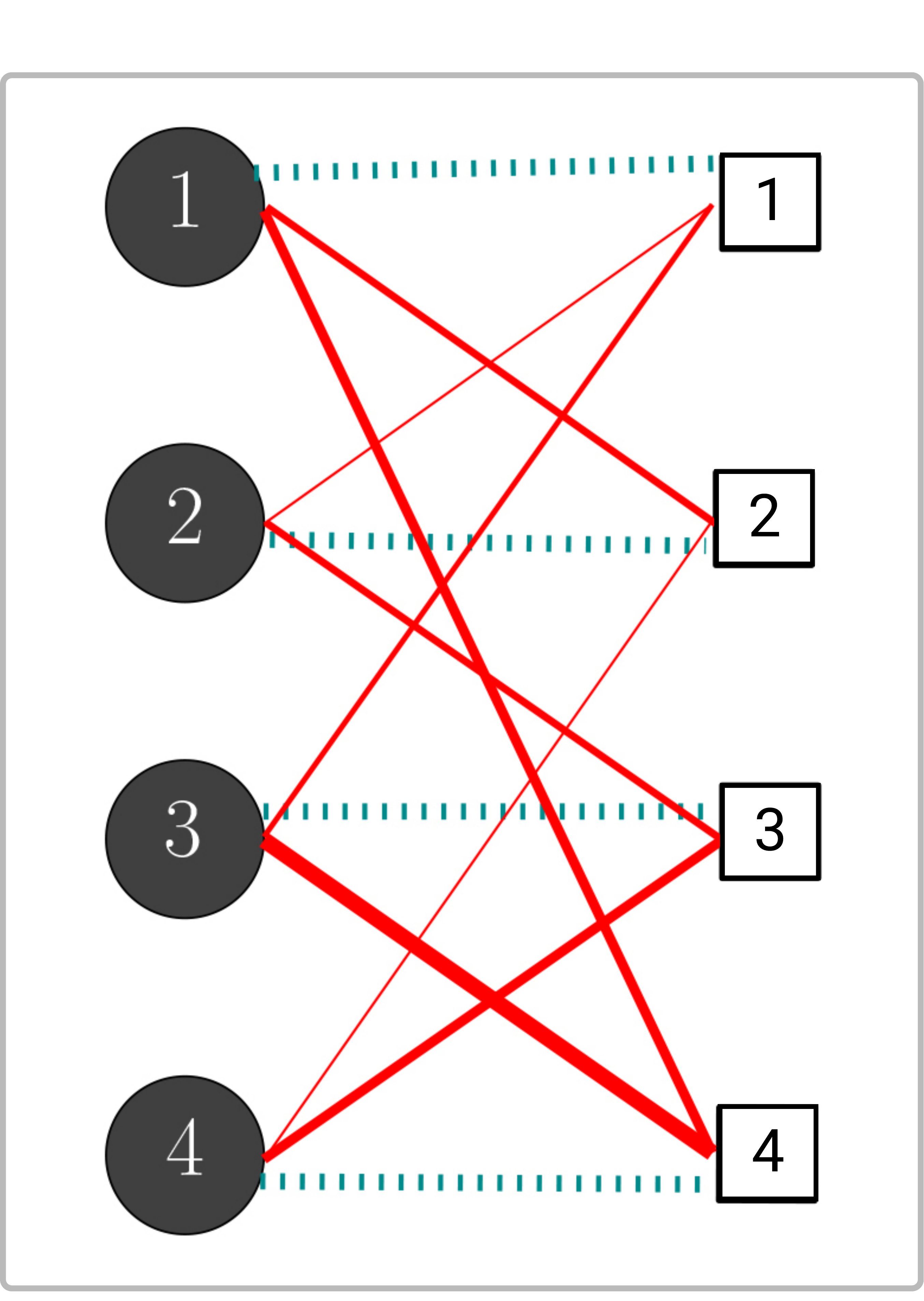}  &
      \includegraphics[width=0.22\textwidth]{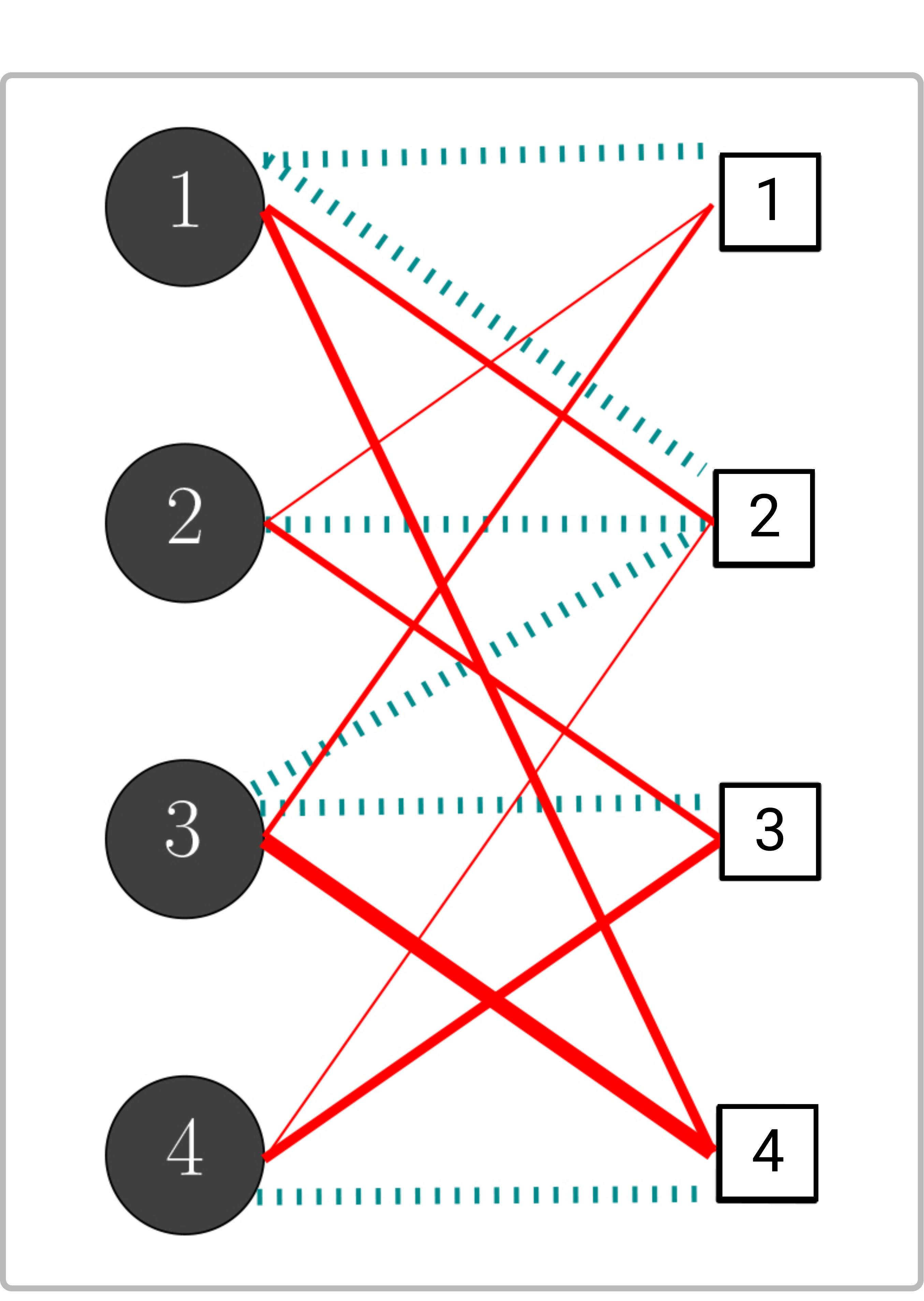}  &
      \includegraphics[width=0.22\textwidth]{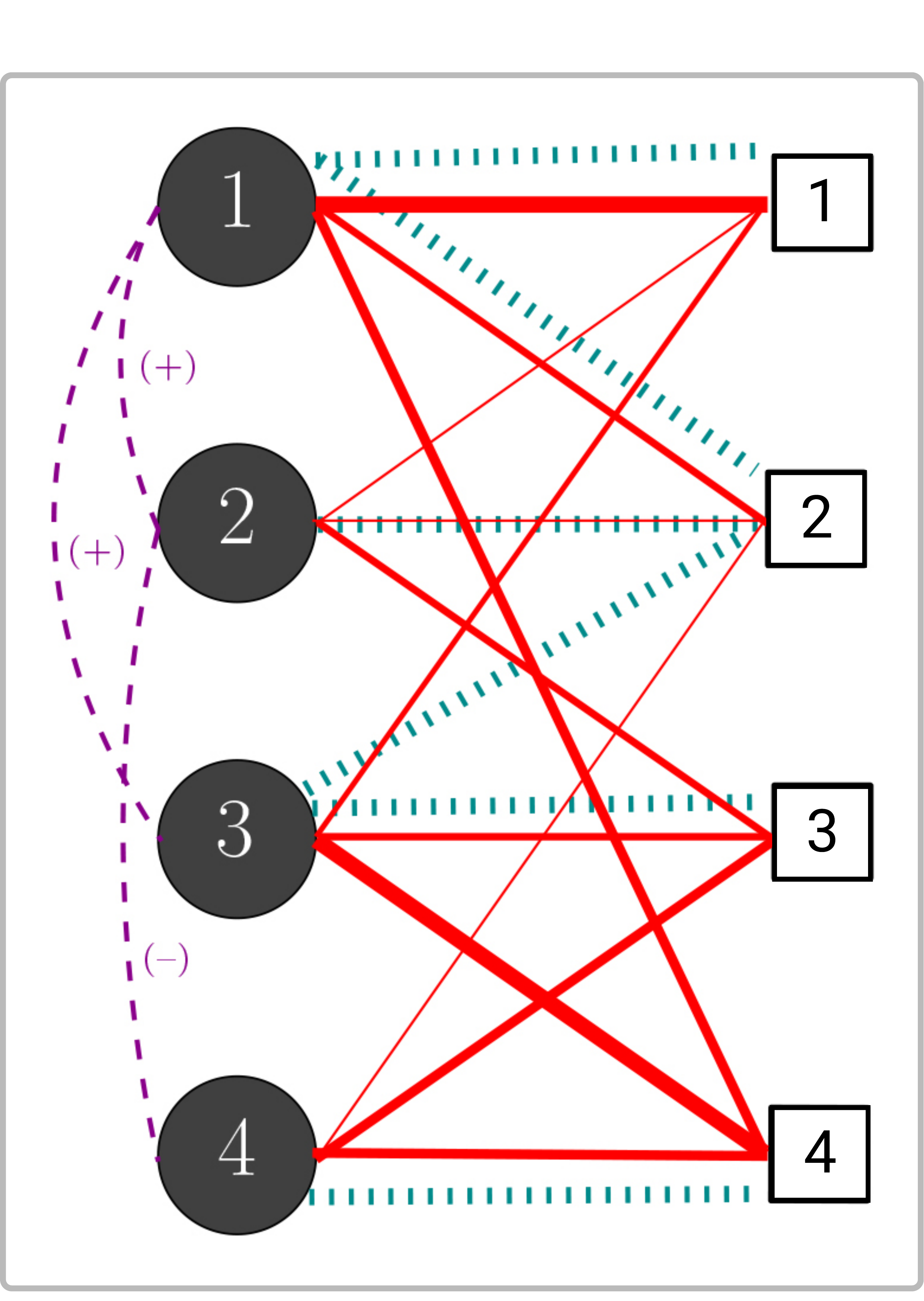} &
      \includegraphics[width=0.22\textwidth]{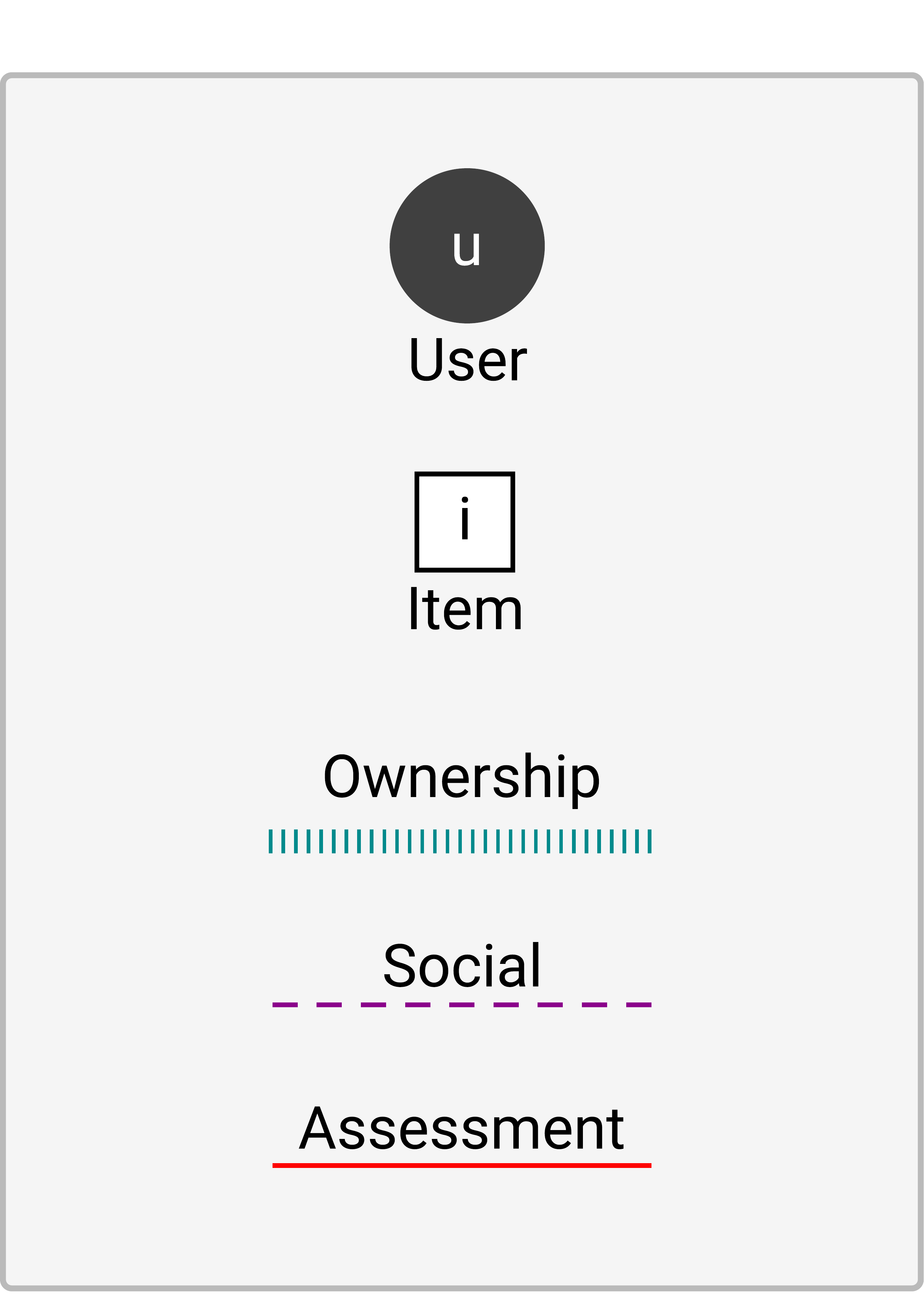} \\
      (e) \small Peer Assessment, Solo Ownership & (f) \small Peer Assessment, Group Ownership 
      &(g) \small Assessment, Group, Social Net 
      &(h) \small Legends
    \end{tabular}
    \caption{Different instantiations of Social-Ownership-Assessment Network (SOAN): (a) assessments provided by users to items as weighted edges; thicker lines represents higher edge weights; (b) self assessments of users for their own item; (c) combination of self and peer assessments of individual contributions; (d) combination of self and peer assessment of both individual and group contributions; (e) peer assessments of individual contributions; and (g) combination of self and peer assessments of group contribution with the presence of social networks between users for capturing conflict of interest.}
    \label{fig:graph_structure}
\end{figure*}

The user-item assessments can be represented by \emph{assessment matrix} $\mathbf{A} = [A_{ui}]$, where $A_{ui}$ is the assessment (e.g., grade or rating) of user $u \in \calU$ for item $i \in \calI$. We let $A_{ui}=0$ when the user $u$'s assessment for item $i$ is missing; otherwise $A_{ui}\in 
\mathbb{R}^+$. As the assessment matrix $\mathbf{A}$ is sparse, we equivalently represent it by an undirected weighted bipartite graph, consisting of two different node types of users $\calU$ and items $\calI$, and weighted assessment edges between them (see Figure \ref{fig:graph_structure}a as an example).

We introduce a \emph{social-ownership-assessment network (SOAN)}, an undirected weighted multigraph, consisting of three types of social, ownership, and assessment relationships on two node types of users and items. In addition to the assessment matrix $\mathbf{A}$, this network consists of two other adjacency matrices: \emph{social matrix} $\mathbf{S} = [S_{uv}] \in \mathbb{R}^{n\times n}$ and \emph{ownership matrix} $\mathbf{O} = [O_{ui}] \in \mathbb{R}^{n\times m}$. The social matrix $\mathbf{S}$, by capturing the friendship and foe relationships between users $\calU$, can accommodate ``conflict of interest'' information. The ownership matrix $\mathbf{O}$, by capturing which users to what extent own or contributed to an item, not only completes conflict of interest information but also provides flexibility of modeling group contributions, self-evaluation, etc. We let $G = (\mathbf{S}, \mathbf{O}, \mathbf{A})$ denote the tuple of all three networks of SOAN.  Figure \ref{fig:graph_structure} demonstrates some instantiations of our models for various settings. 
%
%
Our SOAN model offers important advantages over the existing peer assessment models (e.g., \cite{peerRank14,tunedmodel,fang2017rankwithta, vancouver}):

\vskip 2mm
\noindent \textbf{Expressiveness.} Our model is more \emph{expressive} as it facilitates the representation of many various peer assessment settings that could not be accommodated in the existing models. Its expressive power can be realized in the settings such as self assessments (Figure \ref{fig:graph_structure}b), peer assessments for both solo and group work (Figures \ref{fig:graph_structure}e and \ref{fig:graph_structure}f), and the mixtures of peer and self assessments for solo and group work (Figures \ref{fig:graph_structure}c and \ref{fig:graph_structure}d). For all of these settings, our SOAN model can also express conflict of interest (which is neglected in other models) through a social network (see Figure \ref{fig:graph_structure}g).   

\vskip 2mm
\noindent \textbf{Less Assumptions.} Dissimilar to some existing models (e.g., \cite{peerRank14,tunedmodel,fang2017rankwithta,wang2019sspa}), our model avoids making explicit or implicit assumptions about the relationships between ground-truth values (or grades) and the quality of peer assessments. However, it is still flexible enough to learn such correlations from assessment data if it exists. Our experiments below have shown that our model outperforms other models with restrictive assumptions regardless of whether their assumptions are present in the data or not.    



\subsection{Graph Convolution Networks}
Our learning task is semi-supervised. Given a social-ownership-assessment network $G = (\mathbf{S},\mathbf{O},\mathbf{A})$ and a set of ground-truth valuations $V_{\mathcal{D}} = \{v_j | j \in \mathcal{D}$\} for a subset of items $\mathcal{D} \subset \calI$ , we aim to predict $v_i$ for $i \notin \mathcal{D}$. More specifically, we aim to learn the function $f(i|\bftheta, G)$ for predicting the ground-truth valuation $v_i$ by $\hat{v}_i=f(i|\bftheta, G)$.  The model parameters $\bftheta$ are learned from both the observed ground-truth valuations $V_{\mathcal{D}} = \{v_j | j \in \mathcal{D}$\} and social-ownership-assessment network $G$. We formulate the function $f$ by a modified graph convolution network (GCN) with a logistic head:
\begin{equation}
f(i|\bftheta, G) = \sigma\left(\mathbf{w}^{(o)}\mathbf{z}_i+b^{(o)}\right),
\label{eq:soan-reg}
\end{equation}

where $\sigma(.)$ is the sigmoid function for converting the linear transformation of the node $i$'s embedding  $\mathbf{z}_i$ into its predicted valuations. Here, $\mathbf{w}_o$ and $b_o$ are the weight vector and the bias parameter for the output layer. The node (i.e., item) embedding $\mathbf{z}_i$ is computed with $K$ layers of graph convolution network. Let $\mathbf{H}^{(l)}$ be the $(n+m)\times d$ matrix of $d$-dimensional node embeddings at layer $l$ for all users $\calU$ and items $\calI$ such that user $u$ and item $i$'s vector embeddings are located at the $u$-th and $(m+i)$-th rows, respectively. In Eq.~\ref{eq:soan-reg}, the item $i$'s embedding  $\mathbf{z}_i$ is the $(m+i)$-th row of $\mathbf{H}^{(K)}$ with the updating rule of 
\begin{equation}
\mathbf{H}^{(l+1)} = g^{(l)}\left(\mathbf{D}^{-1}\mathbf{M}\mathbf{H}^{(l)}\mathbf{W}^{(l)}\right).
\label{eq:gcn-soan}
\end{equation}

The matrix $\mathbf{M}$ is constructed from social-ownership-assessment graph $G=(\mathbf{S}, \mathbf{O}, \mathbf{A})$ by   $\mathbf{M}= 
\begingroup 
\setlength\arraycolsep{2pt} 
\begin{pmatrix*}[l] 
\mathbf{S}&\mathbf{P}\\ 
\mathbf{P}^\top&\mathbf{0}_m\\ 
\end{pmatrix*} + \mathbf{I}
\endgroup$, where $\mathbf{P} = \mathbf{O}+\mathbf{A}$, $\top$ is the transpose operator, $\mathbf{0}_m$ is $m\times m$ zero matrix, and $\mathbf{I}$ is the identity matrix. In Eq.~\ref{eq:gcn-soan}, $\mathbf{D}$ is the diagonal matrix with $D_{ii} = \sum_j \ind{M_{ij} \neq 0}$ with $\ind{.}$ as the indicator function.\footnote{One can view $D_{ii}$ as the unweighted degree of the node $i$'s in the social-ownership-assessment network extended with self-loops on all nodes.} The core idea in Eq.~\ref{eq:gcn-soan} is to update the node embeddings at layer $l+1$, denoted by $\mathbf{H}^{(l+1)}$, from layer $l$'s node embeddings $\mathbf{H}^{(l)}$. This update includes the multiplication of the layer $l$'s embeddings $\mathbf{H}^{(l)}$ by the normalized matrix $\mathbf{D}^{-1}\mathbf{M}$,  then linear transformation by learned weight matrix $\mathbf{W}^{(l)}$ at layer $l$, and finally passing through a non-linear activation function $ g^{(l)}$. The initial embedding matrix  $\mathbf{H}^{(0)}$ can be node-level features (e.g., textual features for items, user profiles for users, etc.). When the node-level features are absent, the common practice is to initialize the embeddings with the one-hot indicators \cite{GCN17, wang2020unifying, GRLbook}. As our GCN is built upon our SOAN model, we refer to this combination as \emph{GCN-SOAN}. 

The updating rule in GCN-SOAN (see Eq.~\ref{eq:gcn-soan}) benefits from row normalization of the adjacency matrix similar to many other graph neural networks \cite{wang2019demystifying, GAT, wang2020unifying, wang2018attack}. As the choice of an effective normalization technique is an application-specific question \cite{GRLbook, chen2020learning}, we have decided to normalize our weighted SOAN model by taking an unweighted average, which has been suggested as a solution to address the sensitivity to node degrees for neighborhood normalization \cite{GRLbook}.


GCN can aggregate the information from multi-hop neighborhoods (e.g., neighbors, neighbors of neighbors, and so on) of SOAN. This overarching aggregation makes GCN-SOAN well-equipped to capture the assessment behaviors and patterns. For example, in our experiments below, we demonstrate how GCN-SOAN learns strategic behaviors in peer assessment and predicts the ground-truth valuations more accurately (compared to other methods).

\vskip 2mm
\noindent \textbf{Learning.} Given the social-ownership-assessment network $G$ and a small training set of ground-truth valuations $V_\mathcal{D}$, one can learn GCN-SOAN parameters by minimizing the \emph{mean square error} for its predictions: 
$$
L(\bftheta|G, \mathcal{D}) = \frac{1}{|\mathcal{D}|}\sum_{i=1}^{|\mathcal{D}|}\big(v_i-f\left(i|\bftheta,G\right)\big)^2,
$$
where $|\mathcal{D}|$ is the number of items in the training dataset, and $f\left(i|\bftheta,G\right)$ is the estimated valuation of GCN-SOAN for item $i$. This loss function can be minimized by using gradient-based optimization techniques (e.g., stochastic gradient descent, Adam, etc.).

As opposed to many existing peer assessment systems with unsupervised learning approaches (e.g., \cite{tunedmodel, vancouver, peerRank14}), we deliberately have adopted a semi-supervised learning approach for predicting ground-truth assessment. This choice offers many advantages at some cost of access to a small training data. By learning from the training data, GCN-SOAN is well-equipped to mitigate the influence of strategic behaviors, assessment biases, and unreliable assessments in peer assessment systems. Of course, the extent of this mitigation depends on the size of training data. For the deployment, one can control the number of ground-truth valuations in training data to achieve intended prediction accuracy as the elicitation of ground-truth valuations from an expert (e.g., instructor grades an assignment) can be incremental as it needed.

\section{Experiments}
We evaluate our GCN-SOAN model on various datasets while comparing it against other peer assessment methods. We explain in detail our datasets and experimental methodology, and report our results. Our results demonstrate notable improvements compared to the existing baselines.


\subsection{Datasets}\label{sec:data}

We run extensive experiments on real-world and synthetic datasets. While the real-world datasets allow us to assess the practical efficacy of our approach, we generate various synthetic data to assess its robustness and reliability in various settings (e.g., strategic assessments, biased assessments, etc.)

\begin{table*}[tb]
\begin{center}
\begin{tabular}{cccccccccc}
\toprule
 & \multicolumn{3}{c}{Average Grades}&\multicolumn{6}{c}{Number of} \\ 
\cmidrule[0.5pt](lr){2-4} 
\cmidrule[0.5pt](lr){5-10} 
Asst. ID & Ground-truth & Peer & Self & Exercises & Groups & Students & Items & Peer grades & Self grades\\
\midrule
1 & $0.62\pm0.27$ & 0.70 $\pm$ 0.26 & 0.74 $\pm$ 0.22 & 3 & 75 & 183 & 225 & 965 & 469 \\ 
2 & 0.71 $\pm$ 0.24 & 0.76 $\pm$ 0.23 & 0.80 $\pm$ 0.22 & 4 & 77 & 206 & 308 & 1620 & 755 \\ 
3 & 0.69 $\pm$ 0.33 & 0.75 $\pm$ 0.31 & 0.82 $\pm$ 0.26 & 5 & 76 & 193 & 380 & 1889 & 890 \\
4 & 0.59 $\pm$ 0.27 & 0.68 $\pm$ 0.29 & 0.76 $\pm$ 0.24 & 3 & 79 & 191 & 237 & 1133 & 531 \\ 
\bottomrule
\end{tabular}
\caption{The summary statistics of real-world peer grading datasets.}
\label{table:real-data-stats}
\end{center}
\end{table*}

\vskip 1.5mm
\noindent \textbf{Real-world dataset.} The peer grading datasets of 
Sajjadi et al. \cite{sajjadi} includes peer and self grades of 219 students for exercises (i.e., questions) of four assignments and their ground-truth grades.\footnote{The datasets can be found at \url{http://www.tml.cs.uni-tuebingen.de/team/luxburg/code_and_data/peer_grading_data_request_new.php}. The original datasets are for six assignments. However, two of the datasets have ordinal peer gradings, which can not be used in our experiments.}  For each specific assignment, the submissions are group work of 1--3 students, but each student individually has self and peer graded all exercises of two other submissions (in a double-blind setup). We treat all data associated with each assignment as a separate dataset, where all the submitted solutions to its exercises form the item set $\mathcal{A}$ and the user set $\mathcal{U}$ includes all students who have been part of a submission. We also have scaled peer, self, and ground-truth grades to be in the range of $[0,1]$. Table \ref{table:real-data-stats} shows the statistics summary of our datasets.

\vskip 1.5mm
\noindent \textbf{Synthetic datasets.} In order to evaluate our proposed approach in different settings (e.g., strategic assessments, biased assessments, etc.), we generate various synthetic datasets. We discuss different models used for the generation of our synthetic data.  

\vskip 1mm
\noindent \textit{Ground-truth valuation/grade generation.}
For all $i\in\mathcal{A}$, we sample the true valuation $v_i$ from a mixture of two normal distributions $v_i \sim P(\textsc{x};\bfpi, \bfmu, \bfsigma) = \sum_{c=1}^2 \pi_c \mathcal{N}(\textsc{x}; \mu_c, \sigma_c)$, where  $\bfpi = (\pi_1,\pi_2)$, $\bfmu = (\mu_1,\mu_2)$, and $\bfsigma=(\sigma_1,\sigma_2)$, with $\pi_k$, $\mu_k$, and $\sigma_k$ being the mixing probability, mean, and standard deviation of the component $k$.

\vskip 1mm
\noindent \textit{Social network generation.} We consider two general methods for creating social networks between users.  In \emph{Erd\H{o}s-R\'enyi (ER)} random graph $G(n,p)$ model, each pair of $n$ users are connected to each other with the connection probability of $p$. For our \emph{homophily model} $G(\{v_i\}, \mathbf{O}, \tau)$, we connect two users $j$ and $k$ if  $|v_i-v_l| \leq \tau$ when $o_{ji}=1$ and $o_{kl}=1$. This means that we connect two users if the ground-truth valuations of their owned items are close to each other (i.e, at the distance of at most $\tau$).  This method of social network generation intends to preserve homophily on users' submissions' true grades.

\vskip 1mm
\noindent \textit{Ownership network generation.} For all synthetic datasets, we randomly connect each user to just one item (i.e., one-to-one correspondence between users and items). This setup is in favor of existing peer assessment methods (e.g., \cite{tunedmodel,peerRank14,fang2017rankwithta}), which does not support group ownerships of items.   

\vskip 1mm
\noindent \textit{Assessment network (or peer grades) generation.} To generate peer assessment for each item $i \in \mathcal{A}$, we first randomly select a set of $k$ users $N(i) \subset \mathcal{U}$ such that $|N(i)| = k$. Then, for each $u \in N(i)$, we set $u$'s assessment for item $i$, denoted by $A_{ui}$, using one of these two models. The \emph{strategic} model sets $A_{ui}=1$, if the grader $u$ is a friend of the user $j$ who owns the item $i$ (i.e., $s_{uj}.o_{ji}=1$); otherwise $A_{ui}$ comes from a normal distribution with the mean of $v_i$ and standard deviation of $\sigma_H$. This implies that friends collude to peer grade each others with the highest grade, but would be relatively fair and reliable in assessing a ``stranger.'' The \emph{bias-reliability} model draws $A_{ui}$ from the normal distribution $\mathcal{N}(\textsc{x}; \hat{\mu}, \hat{\sigma})$ with the mean $\hat{\mu} = v_i+\alpha$ and $\hat{\sigma} = \sigma_{max}(1-\beta v_l)$, where $v_l$ is the true valuation of item $l$ owned by user $u$, and $\sigma_{max}$ is the maximum possible standard deviation (i.e., unreliability) for peer graders. Here, the \emph{bias} parameter $\alpha \in [-1,1]$ controls the degree of generosity (for $\alpha > 0$) or strictness (for $\alpha < 0$) of the peer grader.\footnote{It is observed that peer graders tend to grade generously and higher than the ground truth \cite{skewness}. This observation can also be confirmed with our statistics in Table \ref{table:real-data-stats}.} The \textit{reliability} parameter $\beta$ controls the extent that the reliability of the grader is correlated with his/her item's grade (i.e, the peer graders with higher grades are more reliable graders).\footnote{The inductive bias of many peer assessment models (e.g., \cite{peerRank14, fang2017rankwithta, tunedmodel}) include the assumption that the grader's reliability is a function of his/her item's grade. Our bias-reliability model allows us to generate synthetic datasets with the presence of this assumption. So we can compare our less-restrictive GCN-SOAN with those models tailored to this specific assumption in such datasets.} 

\subsection{Baselines}
We compare the performance of our GCN-SOAN model with PeerRank \cite{peerRank14}, TunedModels \cite{tunedmodel}, RankwithTA \cite{fang2017rankwithta}, Vancouver \cite{vancouver}, Average and Median. Average and Median (resp.) outputs the average and median (resp.) of peer grades of each item as the predicted evaluation for the item. As PeerRank, TunedModels, and RankwithTA treat users and items interchangeably, they could not be directly applied to our real-world data with individual assessments on group submissions. For these methods, we preprocess our real-world dataset by taking the average of the grades provided by a group's members for a particular group submission as the group assessment for the group submission.\footnote{In our datasets, as most of the graders for each item are from different groups, this prepossessing step is applied only for a small number of groups.} For the PeerRank and TunedModels, we have used the same parameter settings reported by the original papers. The parameters for RankwithTA and Vancouver are selected by grid search with multiple runs, since the parameters for RankwithTA were not reported, and the suggested parameter of Vancouver results in non-competitive performance.\footnote{In our experiments, the precision parameter for Vancouver is set to 0.1. For RankwithTA, we set 0.8 for the parameter controlling the impact of working ability on grading ability and 0.1 for the parameter controlling the impact of grading ability of a user on his final grade.}

\subsection{Experimental setup}
We implement GCN-SOAN based on PyTorch \cite{pytorch}, and PyTorch Geometric \cite{pytorch-geometric}.
\footnote{
The implementation of GCN-SOAN can be obtained from:  https://github.com/naman-ali/GCN-SOAN/}  
For all experiments, we use two GCN-SOAN convolutional layers with an embedding dimension of 64. We use ELU (Exponential Linear Unit) as activation functions of all hidden layers. We train the model for 800 epochs with Adam optimizer \cite{adam} and a learning rate of $0.02$. We initialize the node embeddings with the vectors of ones. We use Monte Carlo cross-validation \cite{montecarlo} with the training-testing splitting ratio of 1:9 (in synthetic data) or 1:4 (in real-world data), implying that just 10\% or 20\% data is used for training and the rest for testing. To make our results even more robust, we run our model on four random splits and report the average of error over those splits. For each random split, we compute the root mean square error (RMSE) over testing data as the prediction error. 

\begin{table*}[tb]
\begin{center}
\begin{tabular}{lllllllll}
\toprule
 & \multicolumn{4}{c}{\textbf{Peer evaluation}}&\multicolumn{4}{c}{\textbf{Peer and self evaluation}} \\ 
\cmidrule[0.5pt](lr){2-5} 
\cmidrule[0.5pt](lr){6-9} 
\textbf{Model} & Asst. 1 & Asst. 2 & Asst. 3 & Asst. 4 & Asst. 1 & Asst. 2 & Asst. 3 & Asst. 4 \\
\midrule
Average&0.1917&0.1712& \cellcolor{black!10}0.1902&0.1989&0.1944&0.1681&0.2023&0.2117\\ 
Median     & 0.1991 $\downa$ & 0.1843 $\downa$ & 0.2047 $\downa$ & 0.2250 $\downa$ & 0.2111 $\downa$ & 0.1750 $\downa$ & 0.2333 $\downa$ & 0.2538 $\downa$ \\ 
PeerRank   & 0.1913 $\upa$   & 0.1762 $\downa$ & 0.2235 $\downa$ & 0.2087 $\downa$ & 0.1888 $\upa$   & 0.1721 $\downa$ & 0.2203 $\downa$ & 0.2168 $\downa$ \\ 
TunedModel & 0.1919 $\downa$ & \cellcolor{black!25}0.1669 $\upa$   & 0.2110 $\downa$ & 0.2161 $\downa$ & 0.2009 $\downa$ & 0.1680 $\upa$   & 0.2111 $\downa$ & 0.2304 $\downa$ \\ 
RankwithTA & 0.1922 $\downa$ & 0.1903 $\downa$ & 0.2183 $\downa$ & \cellcolor{black!25}0.1740 $\upa$   & 0.1884 $\upa$   & 0.1845 $\downa$ & 0.2137 $\downa$ & \cellcolor{black!25}0.1792 $\upa$   \\ 
Vancouver  & \cellcolor{black!10}0.1851 $\upa$   & 0.1688 $\upa$   & 0.1951 $\downa$ & 0.2071 $\downa$ & \cellcolor{black!10}0.1815 $\upa$   & \cellcolor{black!10}0.1672 $\upa$   & \cellcolor{black!10}0.1945 $\upa$   & 0.2101 $\upa$   \\ 
GCN-SOAN (ours)   & \cellcolor{black!25}0.1795 $\upa$   & \cellcolor{black!10}0.1673 $\upa$   & \cellcolor{black!25}0.1869 $\upa$   & \cellcolor{black!10}0.1822 $\upa$   & \cellcolor{black!25}0.1778 $\upa$   & \cellcolor{black!25}0.1621 $\upa$   & \cellcolor{black!25}0.1840 $\upa$   & \cellcolor{black!10}0.1821 $\upa$   \\ 
\bottomrule
\end{tabular}
\end{center}
\caption{Root mean square error of various methods over two classes of real-world datasets. The first and second best are shown with dark and light gray backgrounds, respectively.
$\upa$ and $\downa$ denote better and worse than Average. GCN-SOAN (ours) is the only method that consistently has outperformed Average for all datasets. Results are averaged over three runs.}
\label{tab:baseline_results}
\end{table*}

\subsection{Results}
We report our results on both real-world and synthetic datasets to evaluate our model in various assessment settings.

\vskip 1.5mm
\noindent \textbf{Real-World Datasets.}
To assess the effectiveness of GCN-SOAN in predicting ground-truth valuations, we compare it against the baseline methods on eight real-world datasets.  These datasets differentiate on (i) which assignment dataset is used and (ii) whether both peer and self-grades are used or only peer grades. For GCN-SOAN, we just create an assessment network, thus allowing us to measure how the assessment network alone can improve the predication accuracy.  As shown in Table ~\ref{tab:baseline_results}, our GCN-SOAN model outperforms others in five datasets and ranked second in remaining ones with a small margin. RankwithTA and TunedModel are the only models that slightly outperform GCN-SOAN for those three datasets. Notably, GCN-SOAN is the only model which consistently outperformed the simple Average benchmark.  This observation is consistent with Sajjadi et al.'s findings \cite{sajjadi} on the same dataset that the existing machine learning methods (not including ours) could not improve results over simple baselines. However, the conclusion does not hold anymore as our machine learning GCN-SOAN approach could consistently improve over simple baselines. This improvement mainly arises from the expressive power and generalizability of GCN-SOAN (discussed in Section \ref{sec:GCN-SOAN}).

\begin{figure}[htb]
   \centering
      \includegraphics[width=0.33\textwidth]{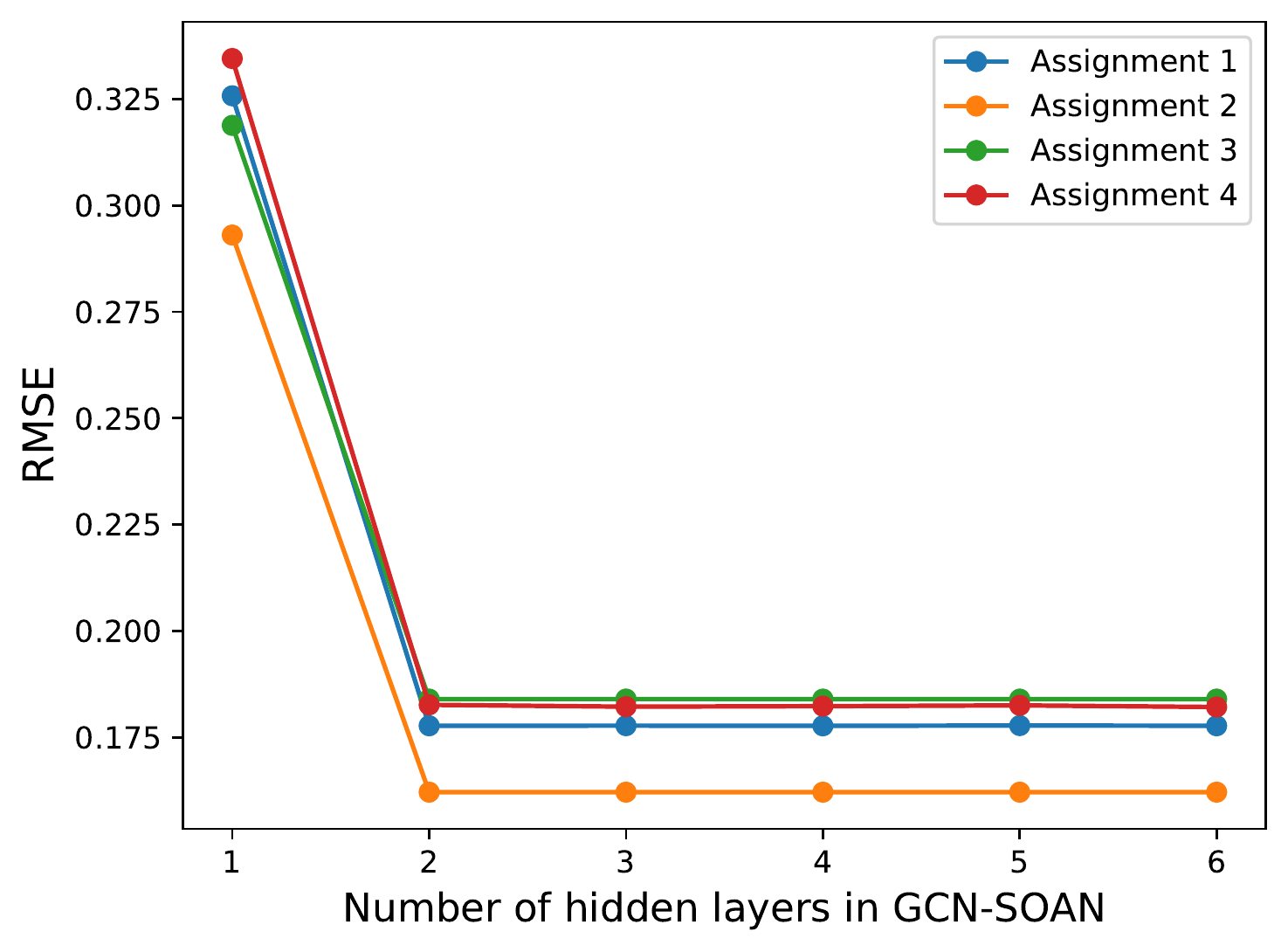}\\
    \caption{Root mean square error for GCN-SOAN, real-world datasets of peer and self evaluation, varying the number of hidden layers. Average over four runs.}
    \label{fig:layers}
\end{figure}

We also further investigate if the efficacy of GCN-SOAN can be improved by adding additional convolutions layers. Figure \ref{fig:layers} shows how the testing error of GCN-SOAN changes by the number of convolutional layers over various datasets.\footnote{The results for peer evaluation datasets were qualitatively similar.} The error first sharply decreases then slightly increases with the number of layers (for all datasets). The optimal error is achieved with two convolution layers. This observation is related to ``over-smoothing'' issues found in GCN-like models \cite{GCN17,sun2021adagcn,xu2018representation,Rong2020DropEdge,luan2019break}. For future work, one can adopt various techniques (e.g., \cite{sun2021adagcn,xu2018representation,Rong2020DropEdge,luan2019break}) to mitigate over-smoothing issue of our GCN-SOAN.   



\begin{figure}
  \centering
  \includegraphics[width=0.33\textwidth]{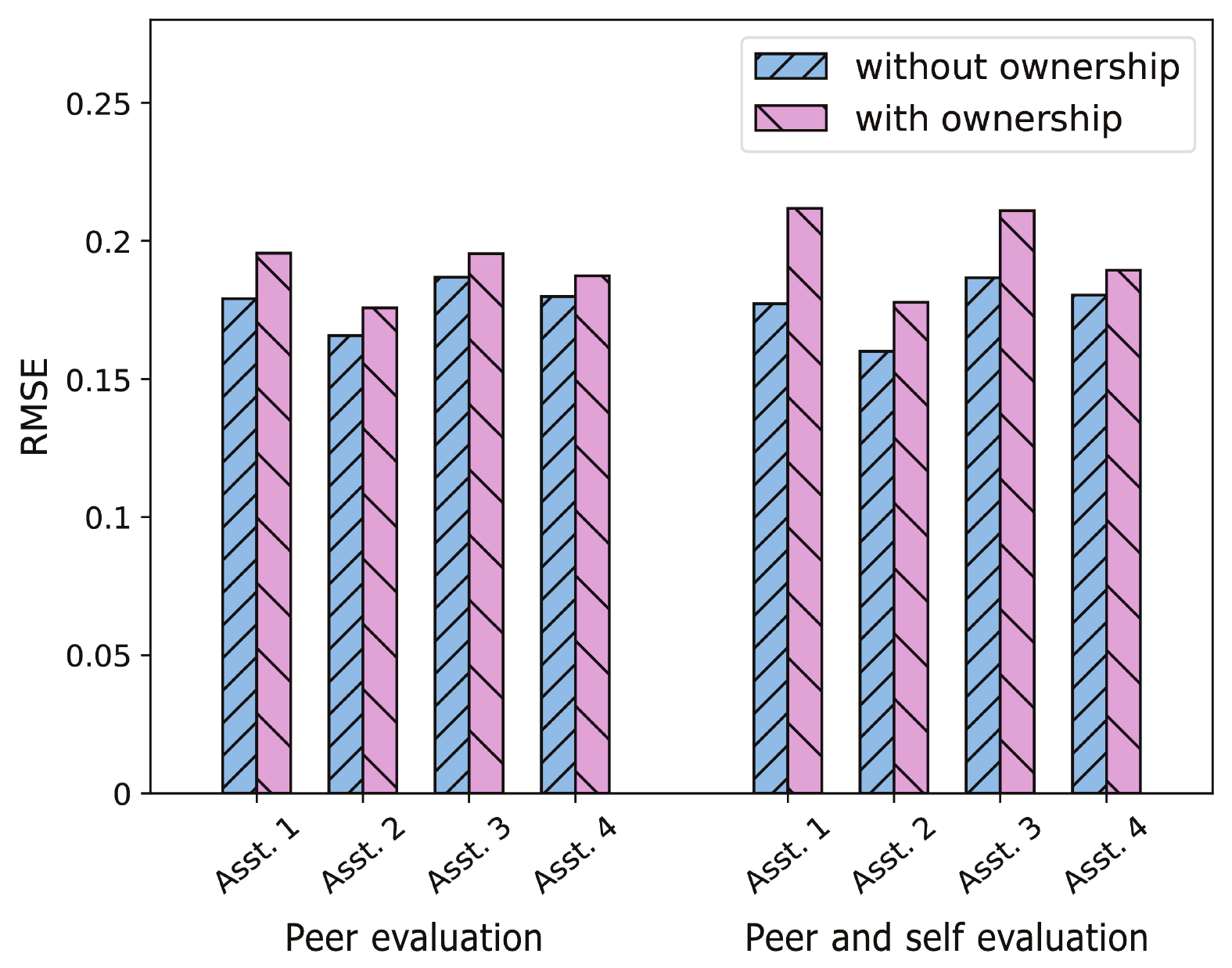}
\caption{Root mean square error for GCN-SOAN for real-world datasets with and without ownership networks. Average over four runs.}
\label{fig:new_experiment_results_real_data}
\end{figure}

We also explore if the addition of the ownership network can further enhance the performance of GCN-SOAN. As shown in Figure \ref{fig:new_experiment_results_real_data}, surprisingly, the addition of ownership network has slightly improved the GCN-SOAN's error. We hypothesis that this downgrade might be partially blamed on how we have linearly combined the ownership network and assessment network in the updating rule of GCN-SOAN (see Eq.~\ref{eq:gcn-soan}). Further algorithmic development to address this shortcoming is an interesting future direction.


\begin{figure*}[htb]
    \centering
    \begin{tabular}{@{\hspace{-2.5pt}}c@{\hspace{-9pt}}c@{\hspace{-9pt}}c@{\hspace{-9pt}}c}
      \includegraphics[width=0.274\textwidth]{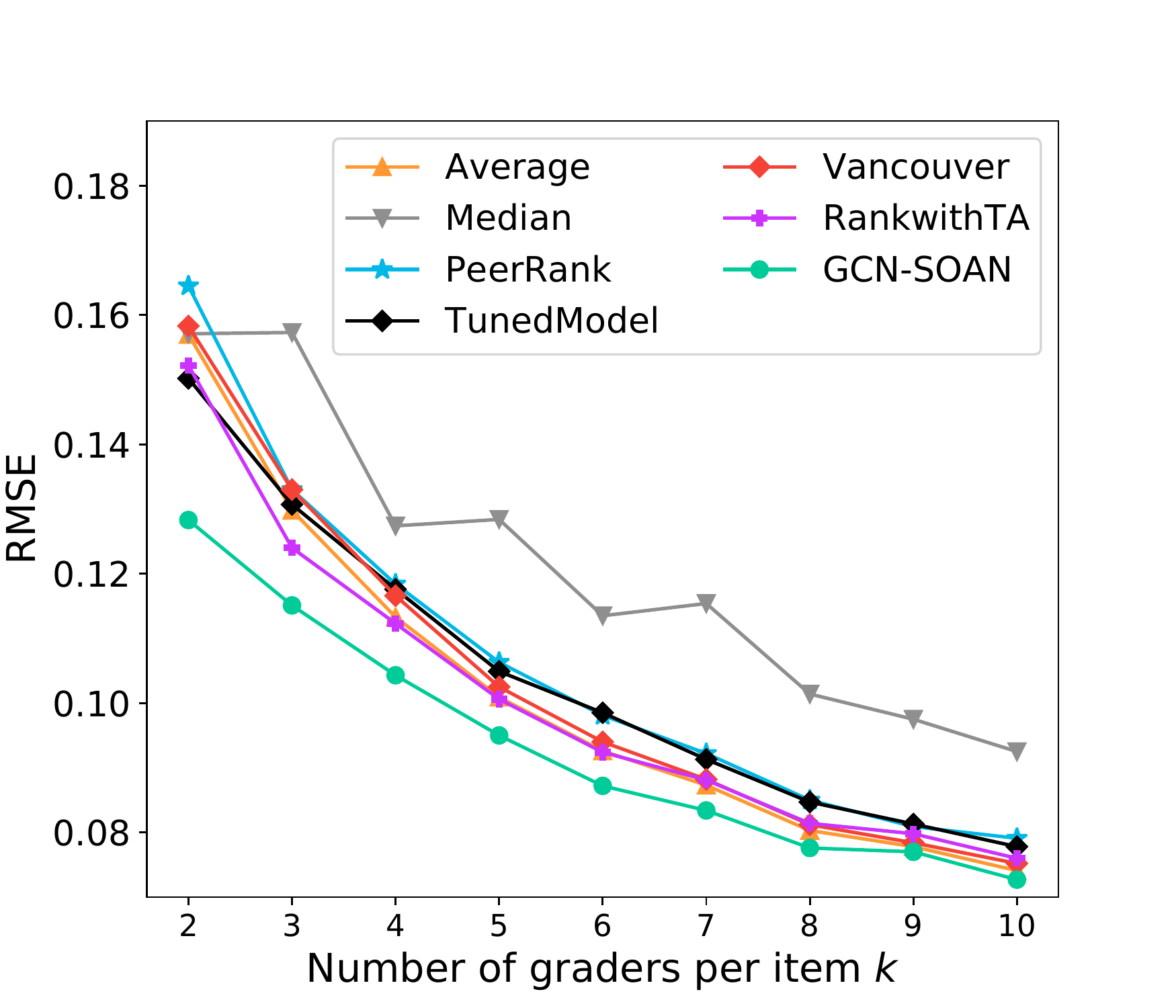}  &
      \includegraphics[width=0.274\textwidth]{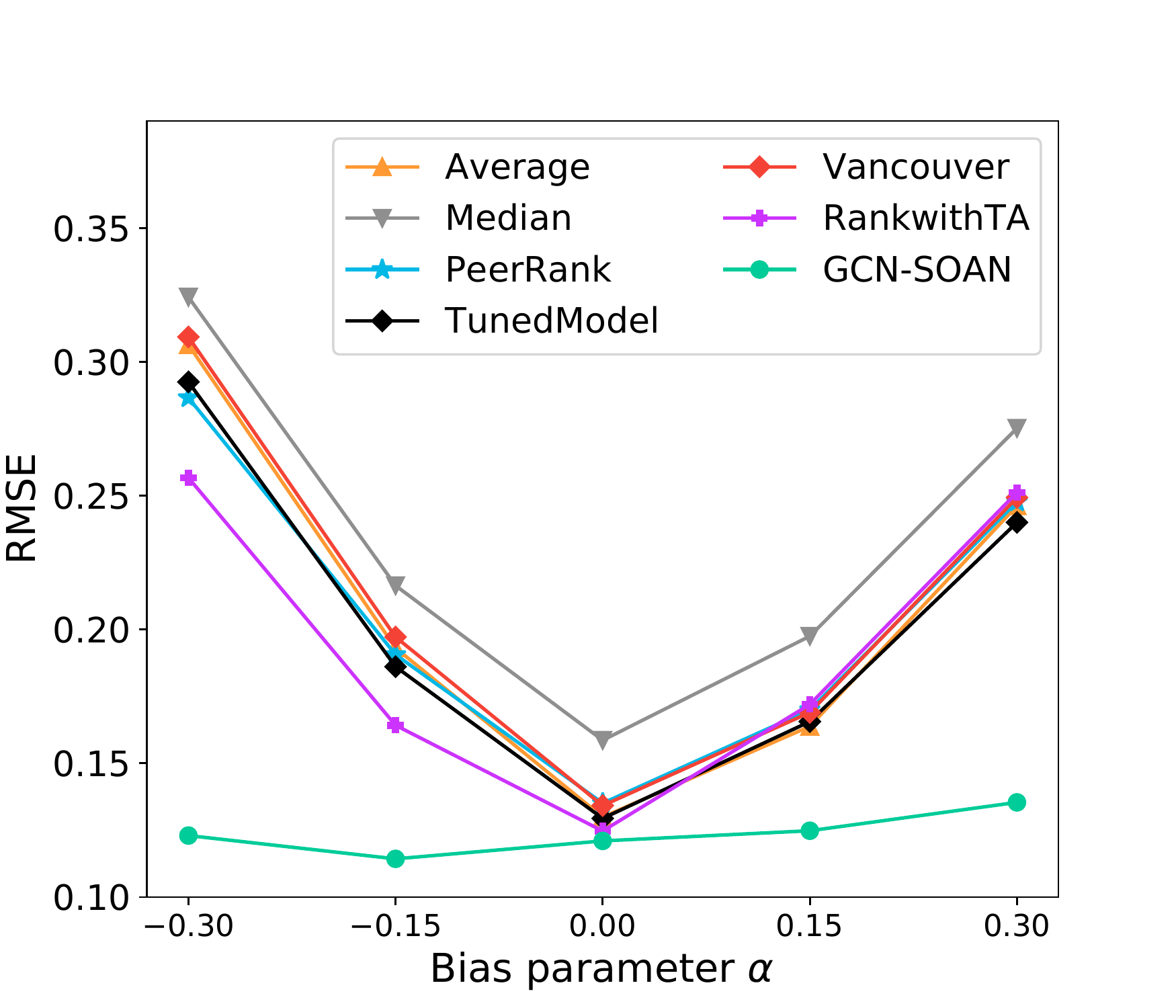} &
      \includegraphics[width=0.274\textwidth]{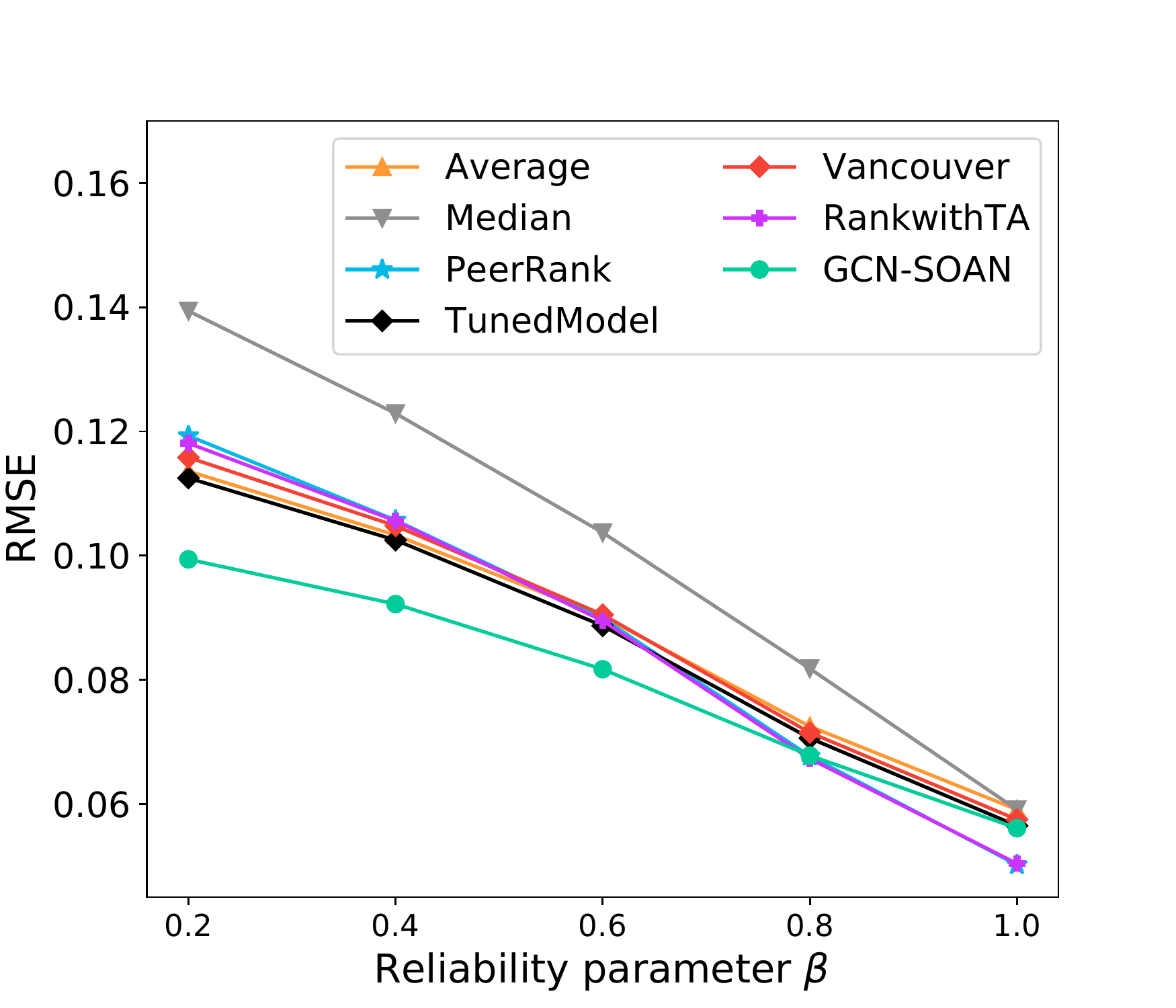} &
      \includegraphics[width=0.274\textwidth]{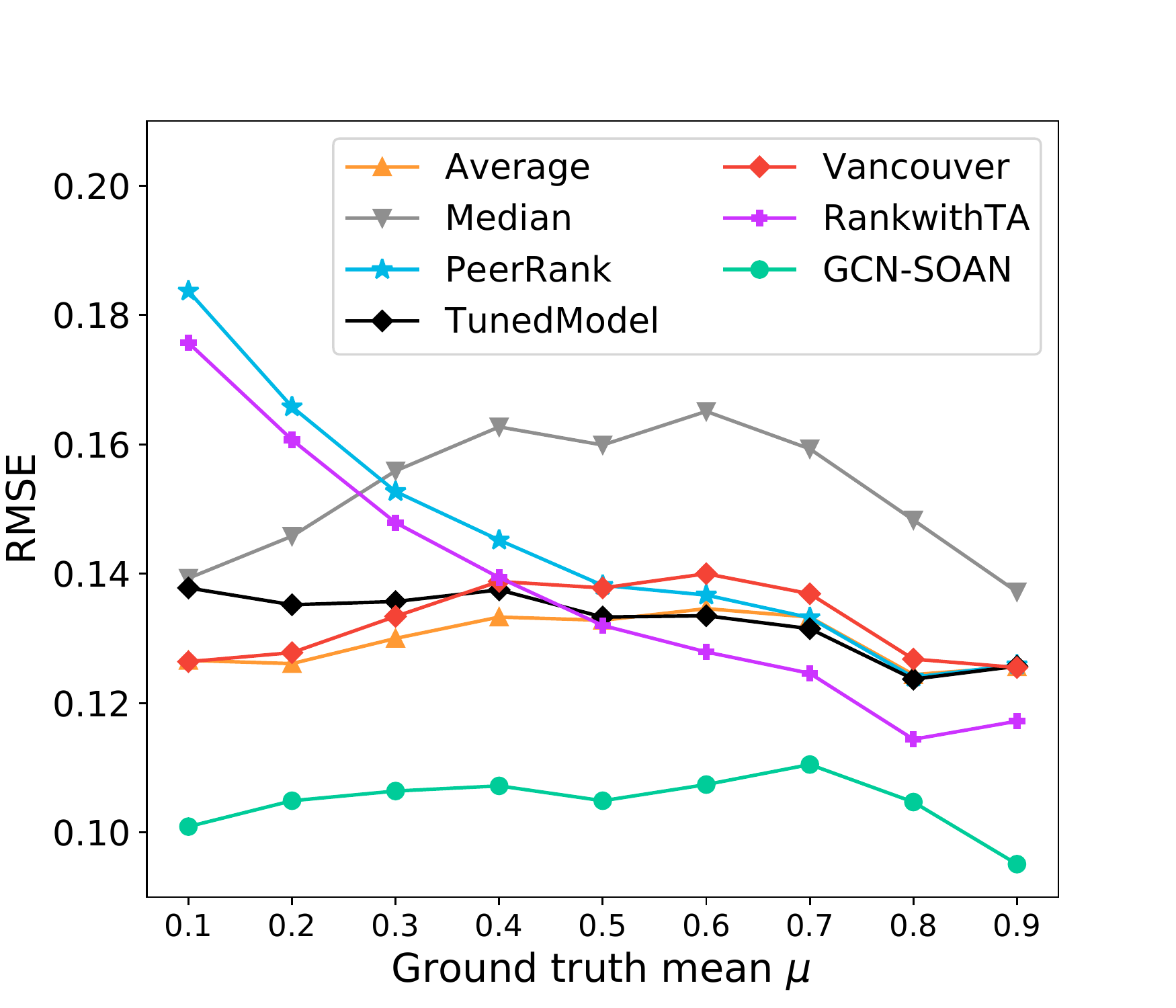} \\
      \small (a) various number of graders $k$  & \small (b) various bias $\alpha$ & \small (c) various reliability $\beta$ &  \small (d) various ground-truth mean $\mu$\\
    \end{tabular}
    \caption{Root mean square errors of various methods, synthetic data with bias-reliability peer generation model, default setting for all parameters but varying (a) number of peer graders $k$, (b) grading bias $\alpha$, (c) reliability parameter $\beta$, or (d) ground-truth mean $\mu$. Average over four runs.}
    \label{fig:reliablity-bias}
\end{figure*}


\vskip 1.5mm
\noindent \textbf{Synthetic Data: Bias-Reliability}.  We run an extensive set of experiments with the bias-reliability peer grade generation model further to assess our GCN-SOAN under various peer assessment settings. For these experiments, we define a \emph{default} setting for all parameter of synthetic generation methods (e.g., bias parameter $\alpha$, reliability parameter $\beta$, etc.). For each experiment, we fix all parameters except one; then, by varying that parameter, we aim to understand its impact on the performance of GCN-SOAN and other baselines. Our default setting includes a number of users $n=500$ and number of items $m=500$; random one-to-one ownership network; $\bfmu=(0.3,0.7)$, $\bfsigma = (0.1,0.1)$, and $\bfpi=(0.2, 0.8)$ for the ground-truth generation method;\footnote{This setting for ground-truth distribution is motivated by two-humped grade distribution in academic classes when most students get a passing grade centered around B grade, but still some students with failing grades centered close the passing-failing border.} the number of peer grades $k=3$,\footnote{This choice of 3 is motivated by the fact that the assessment process is time-consuming and costly in various applications, and most often practical applications do not require more than 3 peer assessment per item (e.g., conference review or peer grading in classrooms).} $\sigma_{max} = 0.25$, bias parameter $\alpha = 0$, and reliability parameter $\beta=0$ for assessment network generation; and no social network generation.\footnote{We have run some other experiments with different default settings (e.g., $n=5000$, different ground-truth distributions, etc.). The results were qualitatively similar and not reported here due to page limit constraints.}

Figure \ref{fig:reliablity-bias}a shows how the prediction error of various methods changes with the number of peer graders $k$ while the other parameters are fixed to the default setting. Unsurprisingly, the performance of all methods improves with $k$. GCN-SOAN not only outperforms others for any $k$, but also exhibits significant improvement over others for a relatively small $k$ (e.g., $k\leq 4$). This superiority of GCN-SOAN with minimal number of peer graders is its strength to make peer assessment suitable and practical for different applications, as so many peer assessment requests will put unnecessary stress and burden on users, thus impeding the practicality of the system. 



Figure \ref{fig:reliablity-bias}b illustrates the errors for each model while changing the bias parameter $\alpha$ (and keeping other parameters fixed to default). GCN-SOAN performs significantly better than other models for any bias values, including generous ($\alpha > 0$) and strict graders ($\alpha < 0$). GCN-SOAN owes this success to its ability to learn students' grading behavior by leveraging a small portion of ground truth grades and assessment network structure. These experiments show that our model could be a great choice for those peer assessment settings where the peer grades are intentionally or unintentionally overestimated/underestimated.

Figure \ref{fig:reliablity-bias}c reports the errors for various values of reliability parameters $\beta$. Recall that the $\beta$ controls the extent that the accuracy of each peer in his/her assessments is correlated with his/her item's grade. Our results show that GCN-SOAN is very competitive to other models, even those built based on this correlation assumption (e.g., \cite{peerRank14, tunedmodel}). We observe that only when $\beta > 0.8$, PeerRank and RankwithTA 
outperform GCN-SOAN. One might argue that $\beta > 0.8$ is implausible scenario in practice. However, this result suggests that our model is still competitive choice for settings in which peer assessment accuracy is correlated with peer success. 


To study how various ground-truth generation distribution impacts the prediction error of various method, we first change the default biomodal mixture of normal distributions (for ground truth generation) to a normal distribution by setting $\mu_1 = \mu_2$ and $\sigma_1=\sigma_2 = 0.15$. Then, we only vary the mean of distribution while other parameters are fixed to default.   
As shown in Figure \ref{fig:reliablity-bias}d, GCN-SOAN consistently outperforms others regardless of the underlying ground-truth distribution. Notably, PeerRank and RankwithTA do not perform well when most users own items with low grades. 


We also explore how homophily (i.e., users' own items with similar ground-truth valuations are more likely to friend with each other) affects the predictive error of peer assessment methods. For this experiment, we generate social networks with our homophily model while varying its $\tau$ parameters. We set other parameters to default. Table \ref{table:homophily} demonstrates that only GCN-SOAN's error decrease with $\tau$. However, the improvements are relatively small. We believe these results highlight an interesting future direction for devising graph neural network solutions, which can capture homophily more effectively.   


\begin{table}[tb]
\begin{tabular}{lllll}
\toprule
\textbf{Model} & $\tau=10^{-4}$ & $\tau=10^{-3}$ & $\tau=10^{-2}$ & $\tau=10^{-1}$ \\
\midrule
Average & 0.1292 & 0.1292 & 0.1292 & 0.1292 \\
Median & 0.1551 & 0.1551 & 0.1551 & 0.1551 \\
PeerRank & 0.1311 & 0.1311 & 0.1311 & 0.1311 \\
RankwithTA & 0.1228 & 0.1228 & 0.1228 & 0.1228 \\
TunedModel & 0.1281 & 0.1281 & 0.1281 & 0.1281  \\
Vancouver & 0.1333 & 0.1333 & 0.1333 & 0.1333 \\
GCN-SOAN (ours) & 0.1212 & 0.1204 & 0.1190 & 0.1189 \\
\bottomrule
\end{tabular}
\caption{Root mean square errors of various methods, synthetic data with bias-reliability peer grade generation model and homophily social network generation model, varying the parameter $\tau$. Average over four runs.}
\label{table:homophily}
\end{table}

\vskip 1.5mm
\noindent \textbf{Synthetic Data: Strategic Assessment}. We study the performance of all peer assessment methods under the strategic model discussed in Section \ref{sec:data}. For this set of experiments, we define this default setting: number of users $n=500$ and number items $m=500$; random one-to-one ownership network; $\bfmu=(0.3,0.7)$, $\bfsigma = (0.1,0.1)$, and $\bfpi=(0.2, 0.8)$ for the ground-truth generation method; the number of peer grades $k=3$ and $\sigma_{H} = 0.25$ for assessment network generation by the strategic model; and ER random graph model with $n=500$ and $p=0.05$ for social network generation.

\begin{figure}
  \centering
  \includegraphics[width=0.4\textwidth]{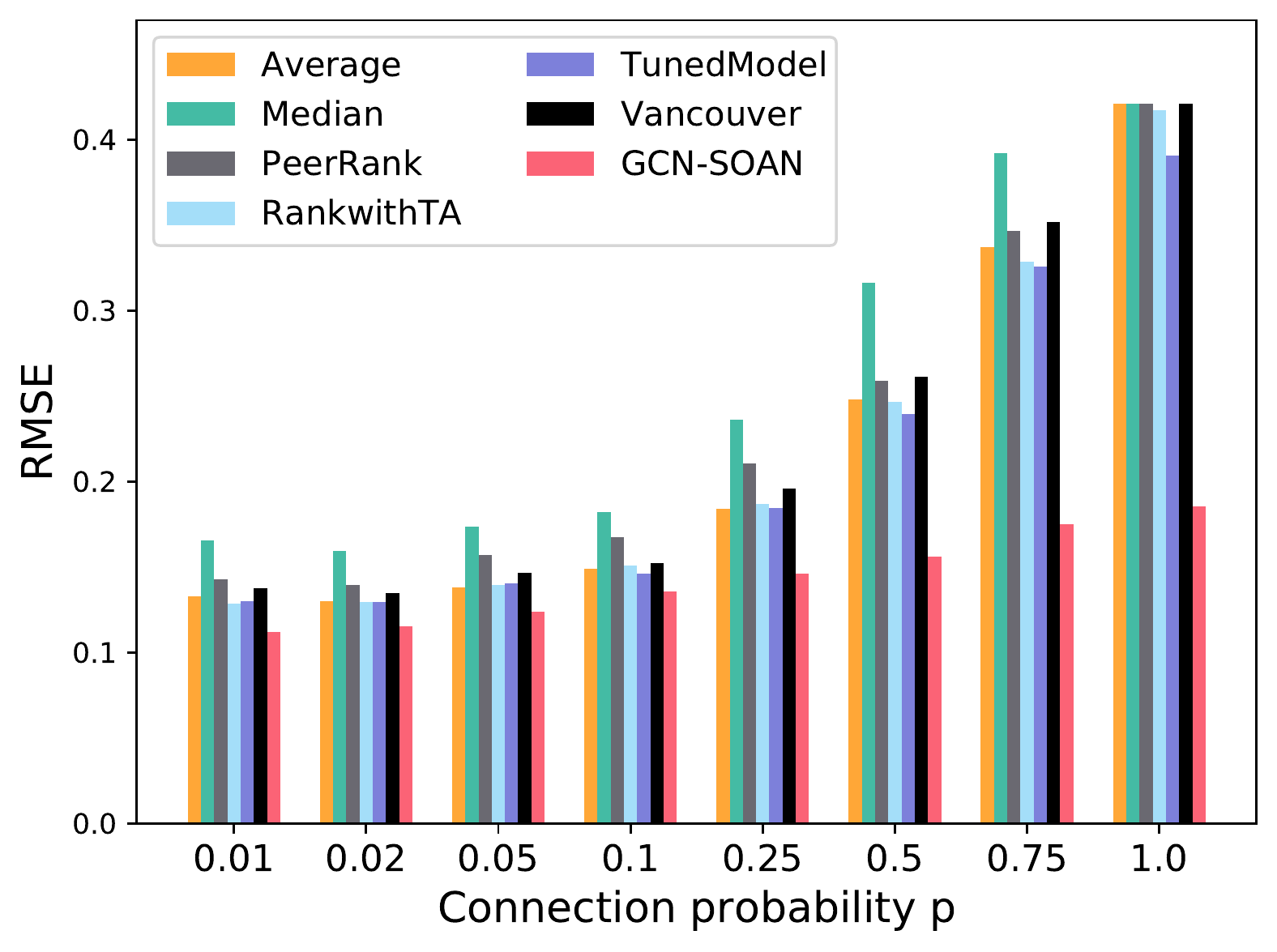}
  \caption{Root mean square errors of various methods, synthetic data with strategic peer grade generation model and Erd\H{o}s-R\'enyi random graph for social networks, varying the connection probability $p$. Average over four runs.}
\label{fig:erdos_study}
\end{figure}

To study how the connection density of colluding social networks impact the accuracy of peer assent methods, we vary the connection probability $p$ while keeping other parameters fixed. Figure \ref{fig:erdos_study} show the outstanding performance of our model compare to other benchmarks and illustrate how our model is more resilient to colluding behaviors. This result suggest that GCN-SOAN is well-eqiupped to detect conflict-of-interest behaviors and mitigate the possible impact of any strategic behaviors. 

\section{Conclusion and Future Directions}
We represent peer assessment data as a weighted multi-relational graph, which we call social-ownership-assessment network (SOAN). Our SOAN can easily express many different  peer assessment setups (e.g., self assessment, peer assessment of group or individual work, etc.). Leveraging SOAN, we introduce a modified graph convolutional network approach, which learns peer assessment behaviors, to more accurately predict ground-truth valuations. Our extensive experiments demonstrate that GCN-SOAN outperforms state-of-the-art baselines in a variety of settings, including strategic behavior, grading biases, etc.

Our SOAN model provides a solid foundation for the broader investigation of graph neural network approaches for peer assessments. Our GCN-SOAN can be extended to mitigate the over-smoothing effect observed in our experiments, or to include a different set of network weights for each relation type of social, assessment, and ownership. Another promising direction is to assess the effectiveness of GCN-SOAN or its extensions on real-world assessment data, with the presence of social network data.

\begin{acks}
We acknowledge the support of Ontario Tech University and the Natural Sciences and Engineering Research Council of Canada (NSERC).
\end{acks}



\bibliographystyle{ACM-Reference-Format} 
\bibliography{main}


\end{document}